\documentclass[a4paper,12pt]{article}
\usepackage[utf8]{inputenc}
\usepackage{graphicx}
\usepackage[hmargin=3cm,vmargin=2cm]{geometry}
\usepackage{setspace}
\usepackage{array}
\usepackage{subfig}
\newcommand\beq{\begin{equation}}
\newcommand\eeq{\end{equation}}
\newcommand\bear{\begin{eqnarray}}
\newcommand\eear{\end{eqnarray}}
\usepackage[fleqn]{amsmath}
\usepackage{authblk}

\title{Engineering the electronic bandgaps and band edge positions in carbon-substituted 2D boron nitride: a first-principles investigation}

\author[1]{Sharmila N. Shirodkar}
\author[1]{Umesh V. Waghmare}
\author[2]{Timothy S. Fisher}
\author[3]{Ricardo Grau-Crespo\thanks{r.grau-crespo@reading.ac.uk}}
\affil[1]{Theoretical Sciences Unit, Jawaharlal Nehru Centre for Advanced Scientific Research, Bangalore 560 064, India}
\affil[2]{School of Mechanical Engineering, Purdue University, West Lafayette, IN 47907-2088, USA}
\affil[3]{Department of Chemistry, University of Reading, Whiteknights, Reading RG6 6AD, UK}

\renewcommand\Authands{ and }

\begin{document}

\maketitle

\begin{abstract}
Modification of graphene to open a robust gap in its electronic spectrum is essential for its use 
in field effect transistors and photochemistry applications. 
Inspired by recent experimental success in the preparation of homogeneous alloys of graphene
and boron nitride (BN), we consider here engineering the electronic structure and bandgap of C$_{2x}$B$_{1-x}$N$_{1-x}$ alloys
via both compositional and configurational modification. We start from the BN end-member,
which already has a large bandgap, and then show that (a) the bandgap can in principle be reduced to about 2 eV
with moderate substitution of C $(x<0.25)$; and (b) the electronic structure of C$_{2x}$B$_{1-x}$N$_{1-x}$ can be further
tuned not only with composition $x$, but also with the configuration adopted by C substituents in the BN matrix.
Our analysis, based on accurate screened hybrid functional calculations, provides a clear understanding of the correlation
found between the bandgap and the level of aggregation of C atoms: the bandgap decreases most when the C atoms are
maximally isolated, and increases with aggregation of C atoms due to the formation of bonding and anti-bonding bands
associated with hybridization of occupied and empty defect states.
We determine the location of valence and conduction band edges relative to vacuum
and discuss the implications on the potential use of 2D C$_{2x}$B$_{1-x}$N$_{1-x}$ alloys in photocatalytic applications.
Finally, we assess the thermodynamic limitations on the formation of these alloys using
a cluster expansion model derived from first-principles.
\end{abstract}

\section{Introduction}
Hexagonal boron nitride (h-BN) has a bulk crystal structure analogous to that of graphite.
An isolated layer of h-BN, with a two-dimensional (2D) honeycomb structure, is thus analogous to graphene\cite{meyer09, nag10}.
But despite the structural equivalence, the electronic properties of these 2D materials differ greatly: graphene (referred to as G here)
is a zero-bandgap semiconductor, whereas a h-BN monolayer (referred to as BN here)
is an insulator with a wide band gap $E_g >$ 5 eV. The simultaneous similarity in crystal structure 
(with a relatively small lattice mismatch of $\sim$1.8\%)
and contrast in electronic behavior, offer potential for a number of exciting applications based on combinations of the two materials.
One current direction of experimental research in this area is the controlled synthesis of in-plane junctions between
conductive G and insulating BN, with the purpose of developing one-atom-thick integrated circuits\cite{sutter, park, liu},
where it is obviously necessary to prevent the unintentional intermixing of the two phases. On the other hand,
one might want to intentionally mix G and BN in order to achieve intermediate electronic bandgaps that could be useful
in electronic or optical devices. However, there are significant thermodynamic constraints for the formation of homogeneous
G-BN alloys as  a strong driving  force exists to segregate G and BN domains/nanophases\cite{lijie}.
Some theoretical work has therefore focused on understanding the effect of domain distribution on the 
electronic structure\cite{chacham, grossman, bhandary, manna, nitesh, peng1} and mechanical properties \cite{peng2, peng3}
of the mixed system. 
In a recent article, Lu {\it et al.} demonstrated the synthesis of highly homogeneous G-BN alloys supported on ruthenium\cite{lu}.
These authors found that the energetics of mixing and demixing processes are modified by the presence of the metal support.
These alloys are still metastable with respect to phase separation,
but their synthesis can be achieved under non-equilibrium conditions at high temperatures,
followed by rapid quenching to prevent the diffusion of species towards segregated domains.
This experimental progress calls for a better theoretical understanding of the properties of highly homogeneous G-BN alloys,
where the formation of domains is inhibited.

In the present work, we study how the electronic structure of 2D 
carbon / boron nitride alloys is determined not only by composition,
but also by the distribution of the ions at a given composition. This is in the spirit of previous research 
which has revealed interesting and potentially tuneable variations of semiconductor bandgaps 
within the configurational space of ion distributions at a fixed composition 
\cite{zunger, seminovski, scanlon, santos}. We start from the BN end-member of the solid solution,
which already has a large gap, and study the behavior of the gap upon C substitution.
In addition to the bandgap, we consider the alignment of the valence and conduction band edges
with respect to the vacuum reference level; this alignment is important in understanding
the electronic behaviour of the interfaces that these alloys form with other metallic or semiconducting materials,
and also the potential of these nanostructures for photochemistry applications.
The compositions considered here can be written as C$_{2x}$B$_{1-x}$N$_{1-x}$, {\it i.e.},
C replaces the same number of B and N atoms, keeping the ratio B/N=1, and low values of $x$ ($x < 1/3$).
These compositions give bandgap values in a range that is useful for optoelectronic and photochemistry applications.

\section{Methods}
For the electronic structure calculations, the 2D materials were simulated using periodic slabs, 
with layers separated by a vacuum gap of fixed width (20 \AA)~along the $c$ direction. 
In the lateral directions, the simulation supercell consisted of 3$\times$3 unit cells and
contained 18 sites (9 B and 9 N sites for pure BN). The cell was substituted with 2, 4 and 6 
C atoms to simulate C$_{2x}$B$_{1-x}$N$_{1-x}$ compositions with $x$= 0.11, 0.22 and 0.33, respectively.
The symmetrically inequivalent configurations at each composition in the 3$\times$3 supercell were 
generated using the Site Occupancy Disorder (SOD) program\cite{sod, sodreview}.
The criterion for two configurations to be symmetrically equivalent is that an isometric transformation converts one into the other,
and the list of possible transformations is obtained from the space group of the parent structure, 
in combination with the supercell translations. 

Our first-principles calculations were based on the density functional theory (DFT) as implemented
in the Vienna Ab initio Simulation Package (VASP)\cite{vasp1,vasp2,vasp3,vasp4}.
The projector augmented wave (PAW) method\cite{paw1,paw2} was used to describe the interaction between 
ionic cores (including the 1$s$ level on each atom) and valence electrons.
An energy cutoff of 520 eV was used for the plane wave basis set expansion.
Integrations in the $k$-space were made using a 8$\times$8$\times$1 uniform mesh of points within
the reciprocal lattice of the supercell (which corresponds to a 24$\times$24$\times$1 mesh in the reciprocal lattice of the unit cell).
The atomic positions and lateral lattice parameters (the $c$ parameter was kept constant)
were optimized with the Perdew–Burke–Ernzerhof (PBE)\cite{pbe} exchange correlation functional,
which is based on the generalized gradient approximated (GGA). At the final geometries, single-point calculations 
based on the screened hybrid functional of Heyd, Scuseria and Ernzerhof (HSE06)\cite{hse03, hse06}
were performed to obtain the electronic structure.
The HSE06 functional is known to yield accurate predictions of electronic bandgaps in semiconductors\cite{heyd},
in contrast to the typical underestimation resulting from GGA functionals.

The evaluation of configuration energies in a larger supercell for the thermodynamic analysis were performed using 
a cluster expansion model\cite{woolverton}, including both nearest and next nearest neighbour clusters for pairs and triplets.
The interaction parameters were fitted to DFT energies in the smaller cell.

\section{Results and discussion}
\subsection{Bandgaps}
\begin{figure}[!h]
\begin{center}
\includegraphics[width=8.3cm]{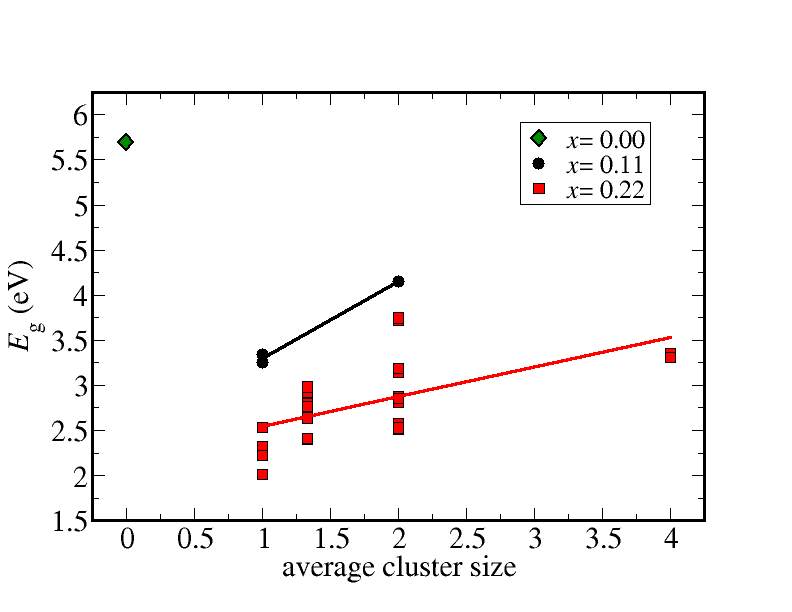}
\caption{Variation of bandgaps with average cluster sizes in the C$_{2}$B$_{8}$N$_{8}$ 
cell composition ($x$= 0.11; black circles), and in the C$_{4}$B$_{7}$N$_{7}$ cell composition ($x$= 0.22; red squares).
The bandgap decreases with C concentration and increases with the average cluster size in the configuration.
The bandgap of h-BN ($x$= 0; green diamond) is given for reference.}
\label{fig1}
\end{center}
\end{figure}

The supercell with composition C$_{2}$B$_{8}$N$_{8}$ ($x$= 0.11) has only 3 symmetrically different site-occupancy configurations:
one configuration where the two C atoms are in nearest-neighbor sites, forming a C-C bond or \textquoteleft dimer',
and two configurations without C-C bonds. The supercell with composition  C$_{4}$B$_{7}$N$_{7}$ ($x$= 0.22) has
30 different configurations, with varying degrees of C aggregation, ranging from the cases where all four
C atoms form a cluster (tetramer), to cases where each C atom is \textquotedblleft isolated" (no C-C bonds).
Our first observation is that  configurations with higher C content tend to have narrower bandgaps,
as expected from the electronic structures of pure G and BN. However, there are also significant variations in bandgaps amongst
configurations corresponding to the same compositions. We have plotted the bandgap versus
the average size of C clusters in each configuration (refer to Figure \ref{fig1}).
Clearly, structures with higher carbon aggregation tend to have wider band gaps as compared to those with
smaller cluster sizes or isolated C atoms. In the C$_{4}$B$_{7}$N$_{7}$ configuration with the narrowest bandgap
($\approx$ 2 eV) all the four C atoms are isolated.

We note that a range of bandgap values still exist for a given average cluster size,
but this dispersion can also be explained by the C distribution.
For example, in the C$_{4}$B$_{7}$N$_{7}$ ($x$= 0.22) composition, when the average cluster size is 2,
there are two possible types of configurations according to the cluster size distribution:
configurations have either one triplet and one isolated carbon (\textquotedblleft1+3"),
or two dimers of carbon atoms (\textquotedblleft2+2"). We find that the 1+3 configurations give bandgaps
smaller than those of the 2+2 configurations.

\begin{figure}[!h]
\begin{center}
\includegraphics[width=8.3cm]{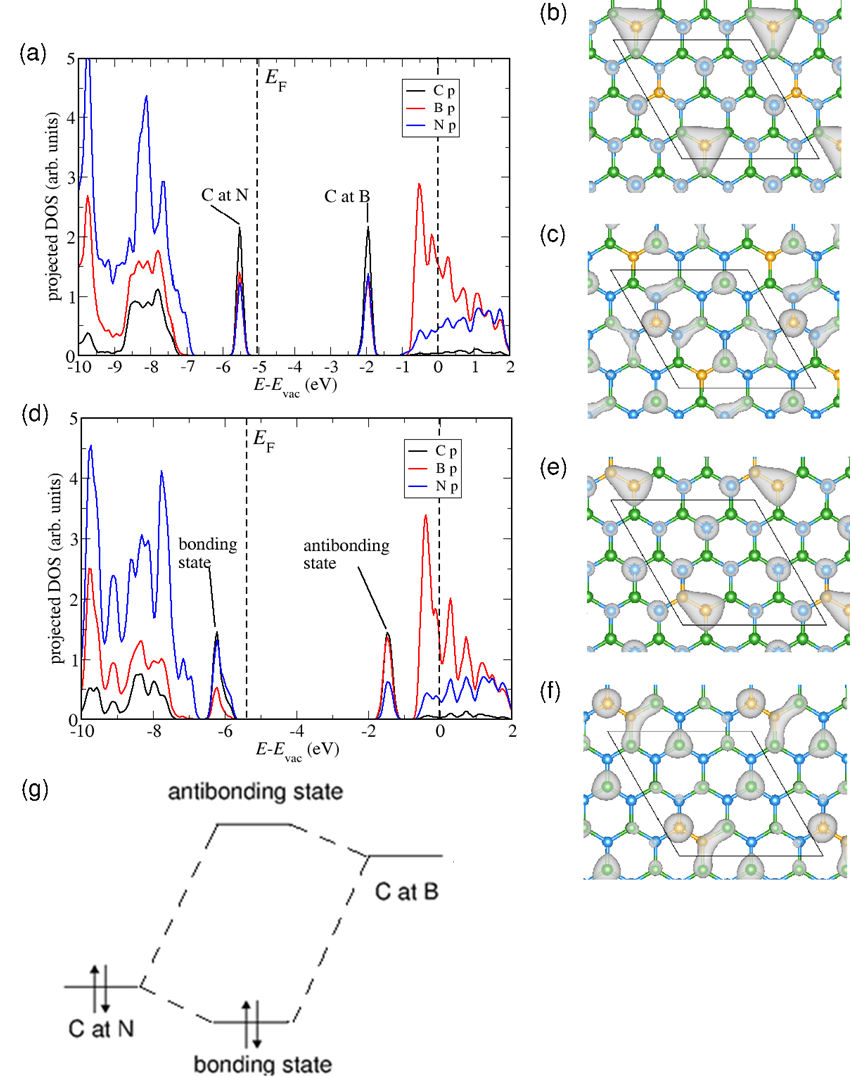}
\caption{Electronic density of states and charge density plot of Highest Occupied Molecular Orbital (HOMO)
and Lowest Unoccupied Molecular Orbital (LUMO) for configurations with isolated (\textquotedblleft 1+1") C atoms
and dimers (\textquotedblleft 2") in C$_{2}$B$_{8}$N$_{8}$ ($x$= 0.22). (a) density of states,
and charge density plots of (b) HOMO and (c) LUMO states of \textquotedblleft 1+1" configuration.
(d) density of states, and charge density plots of (e) HOMO and (f) LUMO states of \textquotedblleft 2" configuration.
(g) Schematic of the formation of bonding and antibonding states from the C/B and C/N defect states in
\textquotedblleft 1+1" configuration. E$_F$ denotes the Fermi energy and the dotted line at 0 is the vacuum level.
Here, C= yellow, B= green and N= blue.}
\label{fig2}
\end{center}
\end{figure}

In order to understand the origin of the trend described above,
we have studied the contribution of the orbitals from different atomic species to the electronic density of states
for two C$_{2}$B$_{8}$N$_{8}$ ($x$=0.11) configurations: one where the C atoms are isolated,
and one where they are forming a C-C dimer (see Figure \ref{fig2}).
We have aligned the electron energies in both cases with respect to the vacuum level,
which was determined from the electron potential in the middle of the vacuum gap in the periodic simulation cell.
The double substitution (C at N and C at B) gives rise to two defect states,
one above the valence band maximum of pure BN and and one below the conduction band minimum of pure BN,
in agreement with a previous report\cite{grossman}.

When the C substituents are separate from each other (Figure \ref{fig2}(a)),
the electron density corresponding to the highest occupied band in the substituted system is clearly localized
at the C atom occupying the N site (Figure \ref{fig2}(b)), whereas the density corresponding to the first empty band
is localized at the C atom occupying the B site (Figure \ref{fig2}(c)).
However, when the two C atoms form a bond, the defect states (Figure \ref{fig2}(d)) arise from the mixing of the orbitals
from the two C atoms. The highest occupied state in this case is the \textquotedblleft bonding" combination (Figure \ref{fig2}(e)),
and the lowest empty state is the \textquotedblleft anti-bonding" combination (Figure \ref{fig2}(f)).
Because the energy difference between the \textquotedblleft bonding" and \textquotedblleft antibonding" states is larger
than that between the isolated C/N and C/B defect states, the bandgap increases when dimerization occurs
(as shown schematically in Figure \ref{fig2}(g)). The formation of a bonding defect state also explains why the 
dimerized configuration is strongly stabilized with respect to the configurations with only C monomers:
the calculated energy difference is 1.66 eV. We will return to the discussion of stabilities below. 

\subsection{Band alignment}
The alignment of the band structure with reference to the vacuum level is important as a basis to understand
the electronic behavior of interfaces involving these 2D materials.
Such referencing also allows us to explore the potential activity in photocatalytic applications,
for example, for hydrogen production via water splitting.
The possibility of making these nanostructures active for photocatalytic hydrogen production is very attractive 
because, if achieved, it could lead to simultaneous hydrogen production and storage in combined C/BN systems. 

Assuming a single-semiconductor cell configuration, and that other materials properties are satisfied,
an ideal photocatalyst for water splitting would have certain important characteristics in its electronic structure.
On the one hand, the positions of the conduction and valence band edges should straddle the redox potentials
for water photolysis\cite{wang, walter, hisatomi}. 
This means that the conduction band edge should be above the energy corresponding to the 
hydrogen evolution reaction (HER), and the valence band edge should be below the energy of the oxygen evolution reaction (OER),
implying also that the bandgap must be wider than 1.23 eV (difference between the HER and the OER).
In fact, once loss mechanisms are accounted for, a bandgap of 2 eV or more is generally necessary\cite{walter}.
On the other hand, the bandgap should not be too wide,
in order to allow the adsorption of photons from the visible part of solar radiation.
It is known that, in the vacuum scale and at pH=0, 
the HER level is located at -4.44 eV, and the OER level is located at -5.67 eV\cite{trasatti}. 
At temperature $T$ and pH$>$0, these energy levels are shifted up by $(k_BT \ln10)\times pH$, 
where $k_B$ is Boltzmann's constant.  
By referencing the electronic levels in our semiconductor solid solutions to the vacuum level
(taken here as the electron potential at the middle of the vacuum gap),
we can assess whether the band edges of the semiconductor are in a favorable position to catalyze the solar splitting of water
under a given set of conditions. 

\begin{figure}[!h]
\begin{center}
\includegraphics[width=7.0cm]{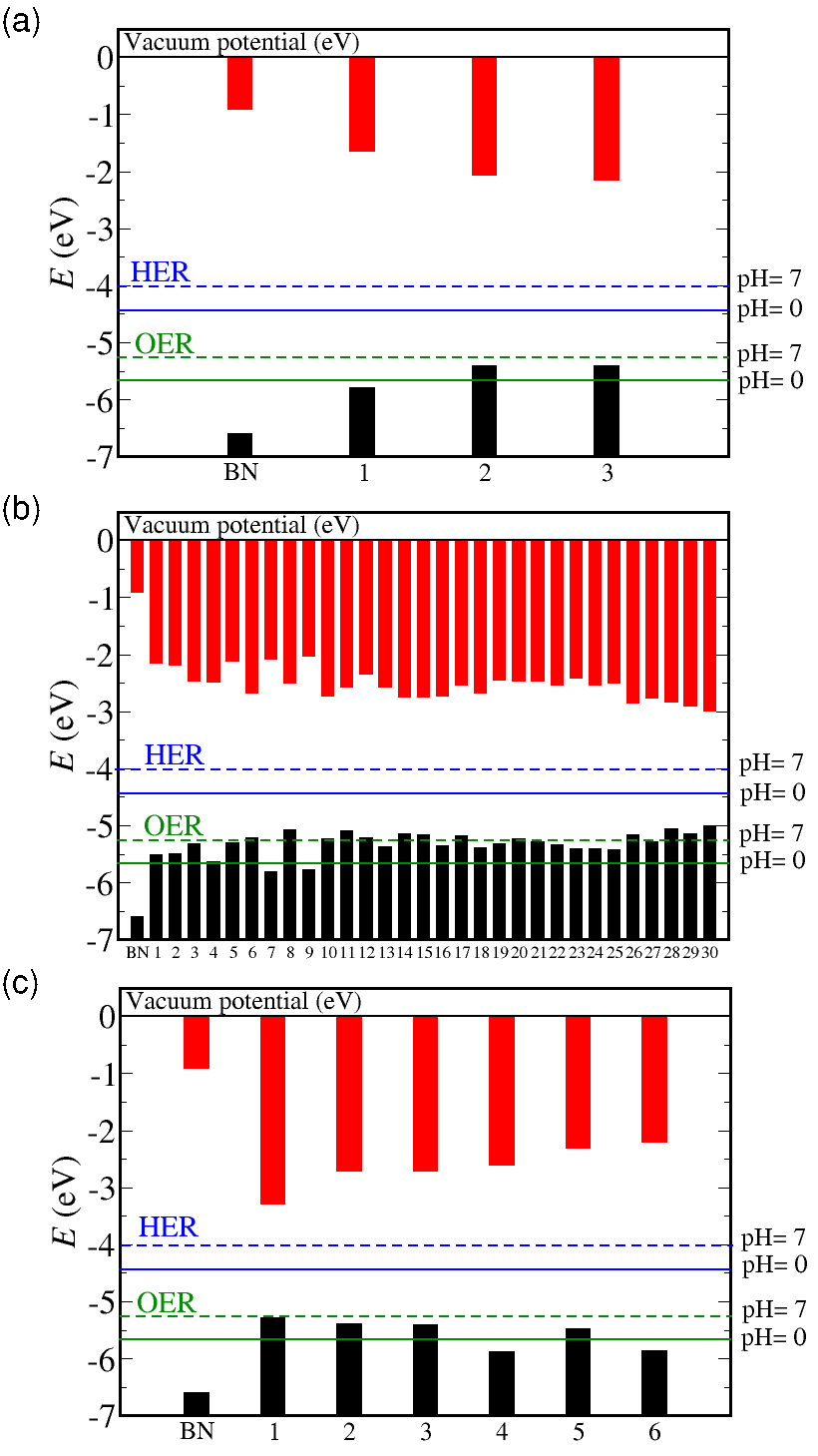}
\caption{Bandgaps and band edge positions calculated with the HSE06 functional for the symmetrically inequivalent configurations
of C$_{2x}$B$_{1-x}$N$_{1-x}$, for (a) C$_{2}$B$_{8}$N$_{8}$ ($x$= 0.11), (b) C$_{4}$B$_{7}$N$_{7}$ ($x$= 0.22),
and (c) for configurations with only dimers in C$_{6}$B$_{6}$N$_{6}$ ($x$= 0.33) with respect to the vacuum potential.
The configurations are arranged on the $x$ axis in increasing order of their total energies. 
The energy levels for the hydrogen evolution reaction (HER) and oxygen evolution reaction (OER) 
are represented by blue and green lines, respectively, both at pH=0 (solid line) and at pH=7 (dashed line)}.
\label{fig3}
\end{center}
\end{figure} 

The positions of the band edges of C$_{2}$B$_{8}$N$_{8}$ ($x$= 0.11) and C$_{4}$B$_{7}$N$_{0.7}$ ($x$= 0.22)
configurations with respect to the vacuum potential and to the HER and OER electrode potentials
at pH=0 and pH=7 are shown in Figure \ref{fig3} (a) and (b) respectively. 
Configurations are sorted in each case by their total energies
(from lower to higher), and the band edge positions of pure BN are given for reference.
Pure BN has has band edges straddling the HER and OER redox potentials
(configurations with this property will be referred to as straddling configurations),
and therefore may exhibit photocatalytic activity, but its bandgap is too wide for efficient solar energy absorption.
For the C$_{2}$B$_{8}$N$_{8}$ supercell composition there is one configuration with 
the two C atoms forming a dimer (configuration 1), and two configurations (2 and 3) with isolated C substitutions 
(no C-C bonds). The dimerized configuration has band edges straddling the HER and OER redox potentials 
at both pH=0 and pH=7, but its bandgap (around 4.1 eV) is still too wide for efficient use of solar energy.
The configurations with isolated C atoms have narrower bandgaps of around 3.3 eV,
but their valence band minimum is too high, above the OER level at pH=0, 
and therefore would not be able to donate a hole for the OER reaction to occur. 
At pH=7 the valence band edges for the isolated C configurations fall just below the OER level, 
but are still too near it to drive this reaction efficiently. 

For the C$_{4}$B$_{7}$N$_{7}$ composition (Figure \ref{fig3}(b)), the 30 symmetrically different configurations 
can be divided into four groups, in order of stability. 
In the first group (configurations 1 and 2 in Figure \ref{fig3}(b)), with the lowest total energies,
the configurations consist of maximally segregated carbon distributions, {\it i.e.} forming C$_4$ clusters.
The bandgap for both these configurations is approximately 3.3 eV, and their valence band maximum 
falls slightly above the OER level at pH=0, and slightly below it at pH=7.
Configurations following these (second group, configurations 3 to 9 in 
Figure \ref{fig3}(b)) are either of \textquotedblleft 2+2" type,  {\it i.e.},
containing two isolated C-C dimers per supercell, or of \textquotedblleft 1+3" type (mixed, {\it i.e.} one triplet and one isolated C).
In particular, for configurations 7 and 9, which are both of \textquotedblleft 2+2" type,
the valence band maximum falls below the OER potential and therefore the conduction and valence band edges
straddle the redox potentials for water photolysis (even at pH=0). 
Their bandgap (around 3.7 eV) is reduced with respect to the dimerized configurations
in the C$_{2}$B$_{8}$N$_{8}$ composition, but is still too wide for efficient photocatalysis.
In the third group (configurations 10 to 26 in  Figure \ref{fig3}(b)),
the configurations are mixtures of \textquotedblleft 1+1+2" and \textquotedblleft 1+3",
and exhibit narrower band gaps. Lastly (fourth part, 27-30), the configurations consist entirely of isolated C substitutions
(\textquotedblleft 1+1+1+1"), and have band gaps in the range of 2.2 to 2.5 eV,
which are suitable for efficient absorption of solar light. But again, the isolated C configurations have a valence band maximum
above the OER level and therefore would not be suitable for a single-semiconductor photolysis cell 
not even at higher pH values (pH=7). However, it is interesting to note that semiconductors with such band positions 
can in principle be utilized as a cathode, in conjunction with other materials (such as TiO$_2$) 
for the anode, in a heterojunction photocatalyst. To be useful as a photocathode, 
the semiconductor band edges are only required to straddle the HER level \cite{wang,walter, hisatomi}. 

Finally, we explore whether we can still achieve configurations straddling both the OER and the HER level
by increasing the C concentration (assuming that such high concentrations can be achieved in a homogeneous solid solution,
which will be discussed in the next section).
We have seen before that only configurations with dimerized substitutions have a valence band edge low enough
to allow hole donation for the OER reaction. Therefore, we have considered all the symmetrically distinct 
\textquotedblleft 2+2+2" configurations with the C$_{6}$B$_{6}$N$_{6}$ ($x$= 0.33) composition.
Our results show that not all these are straddling configurations at pH=0 (Figure \ref{fig3}(c)). 
However, at pH=7, all configurations straddle the OER and HER levels, although the most stable configuration (number 1)
has the valence band maximum too close to the OER level. With the bands positions in the correct region of the energy scale,
it is possible to achieve the desired band alignment if the reaction conditions are changed to slightly higher pH or temperatures.

Therefore, our band alignment results show that a) the band structure of these alloys could be 
engineered to make them work as single-semiconductor photocatalyst for the water splitting reaction
under conditions near to room temperature and neutral pH,
if the C substituents are distributed forming C-C dimers, but these configurations tend to have too wide bandgaps
for efficient solar energy utilization; and b) configurations with only isolated C substituents, 
if successfully synthesized at a reasonably high C concentration, could achieve narrower bandgaps in the desired range
for solar light absortpion but with a valence band edge too high for single-semiconductor water splitting photocatalysts. 
These configurations would satisfy the band edge position requirements for a cathode in a heterojunction photocatalyst.

\subsection{Stability}
\begin{figure}[!h]
\begin{center}
\includegraphics[width=8.3cm]{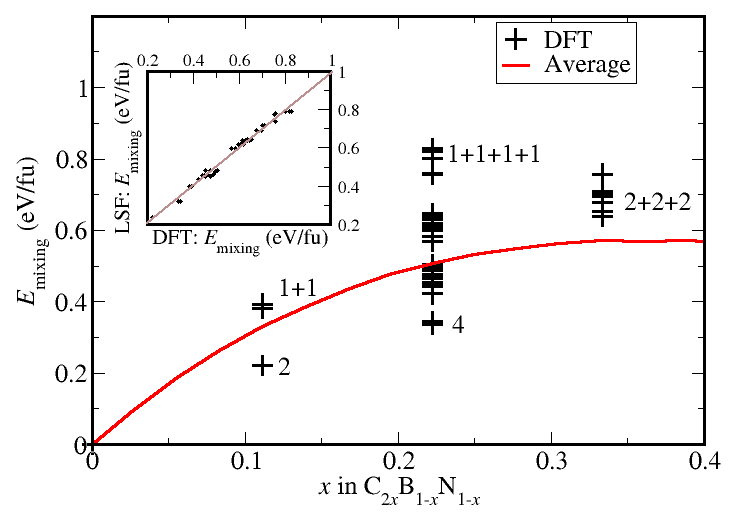}
\caption{Crosses show the DFT mixing energies (per formula unit) for C$_{2x}$B$_{1-x}$N$_{1-x}$ 
configurations and the red line represents the average mixing energies calculated for a converged sample 
of random configurations using a cluster expansion model. 
Inset shows the mixing energies predicted by the model versus DFT for C$_{2x}$B$_{1-x}$N$_{1-x}$.} 
\label{fig4}
\end{center}
\end{figure}

We briefly discuss here the relative energies of the configurations,
and also the stability of the alloy with respect to phase separation.
The calculated mixing energies (energies of the alloy 
with respect to the corresponding amounts of pure BN and graphene) 
are positive and high for all the compositions and configurations, 
confirming that the alloy formation is always a highly endothermic process 
(Figure \ref{fig4}). For reference, we have also plotted the average energy of the  alloy as a function of composition
(seen as red line in Figure \ref{fig4}), as calculated in a larger (6$\times$6) cell for each composition.
This was done using a cluster expansion\cite{woolverton} model with parameters fitted to DFT energies in the smaller cell;
the agreement between the energies calculated at DFT level and those obtained from the cluster expansion is illustrated
in the inset of Figure \ref{fig4}. The configurations evaluated with the cluster expansion and 
included in the average energy all have the C$_{2x}$B$_{1-x}$N$_{1-x}$ stoichiometry. 

The very high mixing energies shown in Figure \ref{fig4} indicate that the free energy of mixing,
despite the stabilizing effect of the configurational entropy, will also be positive at any significant C concentration.
To illustrate this, consider the maximum configurational entropy for this type of alloy, using the ideal expresssion:
\begin{equation}
\label{eq:entropy}
S=-2k_B(x \ln x + (1-x)\ln (1-x)).
\end{equation}
where the factor of 2 comes from the two sublattices with mixed occupancy.
At $x$=0.1 and $T$=1000 K, $-ST$=-56 meV per formula unit, 
which is more than 5 times smaller in absolute value than the mixing energy at that composition. 

We explained above how the formation of C-C bonds lowers the total energy with respect to two isolated C atoms;
we see here that the same trend is observed for larger clusters, which are always energetically favoured over smaller ones.
Thermodynamics clearly drives the system towards the segregation of BN and C domains, and against the formation of an alloy, 
as has been discussed previously by other authors\cite{bhandary, lijie, lu}.
Our results show that isolated carbon substitutions are always the least stable,
and are above the average energy line, particularly for higher concentrations,
which suggest that such configurations might be difficult to achieve.
It is important to note that these alloys must always be synthetized by a non-equilibrium procedure \cite{kalyan,lu}
and therefore it is difficult to predict the extent to which the energy differences between configurations are relevant.
However, our results suggest that in a metastable sample with significant C substitution, obtained by quenching from 
high temperature, isolated C atoms would occur rarely, if at all.
A sample with only isolated C substitutions could only be obtained by such methods at very low C concentrations,
where the configurational entropy dominates the distribution. 
At higher C concentration, a more sophisticated synthesis technique allowing atom-by-atom control would be required. 
On the other hand, configurations with isolated C-C dimers, which are also potentially interesting in photocatalysis, 
might be easier to synthesize because the C-C interaction stabilizes them with respect to isolated C substitutions. 
However, they are still metastable with respect to phase segregation.    

\section{Conclusions}
Our work elucidates the relationship between the distribution of C substituents 
in 2D boron nitride and the resulting electronic structure.
We show that the bandgap decreases with increasing C concentration, as expected, 
and that the reduction in gap is more pronounced the more disperse the distribution of the substituents is.
Hence the smallest bandgaps for a given composition correspond to configurations with only isolated C substituents,
that is, with no C-C bonds. For example, we have found that the substitution of isolated C atoms at a concentration
of only $x$=0.22 brings the bandgap down to 2 eV. 

From the point of view of potential applications in electronics, 
it is important to note that this small gap arises from narrow bands associated with defect states,
leading to large effective masses and low mobilities.
However, for larger concentrations of C substitutions the edge bands become wider.
Therefore when engineering the bandgap of this system one would need to find a balance between gap opening and mobility. 

The analysis of the electronic structure shows that the clustering of the C substituents
leads to mixing of the wavefunctions corresponding to these defect states, forming bonding and anti-bonding levels.
This effect is responsible for the increase in the band gap with C clustering. 

We have also reported the variation of the band edge positions with respect to composition and ion distribution.
We found that configurations consisting of C-C dimers have band alignments 
which are favourable for single-semiconductor 
photocatalysis of the water splitting reaction under conditions near to room temperature and neutral pH,
but these configurations tend to have too wide bandgaps for efficient solar energy utilization. 
On the other hand, configurations with only isolated C atoms have smaller bandgaps. Their band edge
 positions do not straddle the OER level, but satisfy the requirements for a cathode in a heterojunction photocatalyst
(straddling the HER level). However, our thermodynamic analysis shows that configurations with isolated C atoms 
would be very difficult to obtain due to the strong tendency of C substituents to cluster in the BN matrix. 

\section*{Acknowledgements}
The authors thank the UK-India Education and Research Initiative (UKIERI) for funding this collaborative project
through a Trilateral Partnership Grant IND/CONT/2013-14/054. 
Via the UK's HPC Materials Chemistry Consortium, which is funded by EPSRC (EP/L000202),
this work made use of ARCHER, the UK's national high-performance computing service.

\bibliographystyle{prsty}
\bibliography{ref}

\end{document}